\documentclass[10pt, conference, compsocconf]{IEEEtran}
\usepackage{times}
\usepackage{helvet}
\usepackage{courier}

\usepackage[T1]{fontenc}
\usepackage[latin9]{inputenc}
\usepackage{color}
\usepackage{url}
\usepackage{amsthm}
\usepackage{amsmath}
\usepackage{amssymb}
\usepackage{graphicx,threeparttable,diagbox}
\usepackage{footnote,epstopdf,nameref,xspace}

\usepackage{algorithm}
\usepackage{algorithmic}

\usepackage{float}
\usepackage{graphicx}
\usepackage{epsfig}
\usepackage{balance}
\usepackage{multirow}

\graphicspath{{pix/}}
\usepackage{caption}
\usepackage{subcaption}
\ifCLASSINFOpdf
\else
\fi

\usepackage{listings}
\usepackage{xcolor}
\usepackage{color}
\definecolor{keywordcolor}{rgb}{0.8,0.1,0.5}
\definecolor{webgreen}{rgb}{0,.5,0}
\begin{document}
%
% paper titlep
\title{Look into My Eyes: Fine-grained Detection of Face-screen Distance on Smartphones}


%\author{\IEEEauthorblockN{Authors Name/s per 1st Affiliation (Author)}
%\IEEEauthorblockA{line 1 (of Affiliation): dept. name of organization\\
%line 2: name of organization, acronyms acceptable\\
%line 3: City, Country\\
%line 4: Email: name@xyz.com}
%\and
%\IEEEauthorblockN{Authors Name/s per 2nd Affiliation (Author)}
%\IEEEauthorblockA{line 1 (of Affiliation): dept. name of organization\\
%line 2: name of organization, acronyms acceptable\\
%line 3: City, Country\\
%line 4: Email: name@xyz.com}
%}
%\author{
%Zhuqi Li, Weijie Chen, Zhenyi Li and Kaigui Bian\\
%School of EECS, Peking University, Beijing, China\\
%\{lizhuqi,\space chenweijie2013,\space lizhyi,\space bkg\}@pku.edu.cn
%%\\
%}

\author{
\IEEEauthorblockN{Zhuqi Li}
\IEEEauthorblockA{School of Electronic Engineering and Computer Science \\
Computer Science Department\\
Peking University\\
Beijing, China, 100871\\
Email: lizhuqi@pku.edu.cn}
\and

\IEEEauthorblockN{Weijie Chen}
\IEEEauthorblockA{School of Electronic Engineering and Computer Science \\
Computer Science Department\\
Peking University\\
Beijing, China, 100871\\
Email: chenweijie2013@pku.edu.cn}
\and

\IEEEauthorblockN{Zhenyi Li}
\IEEEauthorblockA{School of Electronic Engineering and Computer Science \\
Computer Science Department\\
Peking University\\
Beijing, China, 100871\\
Email: lizhyi@pku.edu.cn}
\and

\IEEEauthorblockN{Kaigui Bian}
\IEEEauthorblockA{School of Electronic Engineering and Computer Science \\
Computer Science Department\\
Peking University\\
Beijing, China, 100871\\
Email: bkg@pku.edu.cn}

}

\maketitle
\begin{abstract}
%\begin{quote}
The detection of face-screen distance on smartphone (i.e., the distance between the user face and the smartphone screen) is of paramount importance for many mobile applications, including dynamic adjustment of screen on-off, screen resolution, screen luminance, font size, with the purposes of power saving, protection of human eyesight, etc. Existing detection techniques for face-screen distance depend on external or internal hardware, e.g., an accessory plug-in sensor (e.g., infrared or ultrasonic sensors) to measure the face-screen distance, a built-in proximity sensor that usually outputs a coarse-grained, two-valued, proximity index (for the purpose of powering on/off the screen), etc.
In this paper, we present a fine-grained detection method, called ``Look Into My Eyes (LIME)'', that utilizes the front camera and inertial accelerometer of the smartphone to estimate the face-screen distance. Specifically, LIME captures the photo of the user's face only when the accelerometer detects certain motion patterns of mobile phones, and then estimates the face-screen distance by looking at the distance between the user's eyes. Besides, LIME is able to take care of the user experience when multiple users are facing the phone screen. The experimental results show that LIME can achieve a mean squared error smaller than 2.4 cm in all of experimented scenarios, and it incurs a small cost on battery life when integrated into an SMS application for enabling dynamic font size by detecting the face-screen distance.
\end{abstract}

\section{Introduction}
The fine-grained detection of face-screen distance on smartphone (i.e., the distance between the user face and the smartphone screen) brings many benefits for mobile utility applications. For better saving the battery life, every smartphone has a built-in proximity sensor that helps turn the screen off (or on) when the detected proximity between the user's face and the phone is small (or large)~\cite{proximitysensor}. Moreover, a phone with a plug-in infrared or ultrasonic sensor is able to determine the face-screen distance, and accordingly adjust the screen resolution for power saving and protecting users' eyesight~\cite{he2015optimizing}, e.g., to increase (or decrease) the screen resolution when the detected face-screen distance is small (or large).

Existing detection techniques for face-screen distance are limited in the following aspects.
\begin{enumerate}
  \item \textbf{Coarse-grained output}. The distance output of the phone's built-in proximity sensor is a coarse-grained and two-valued index, which usually indicates whether the face is close to the screen or not.
  \item \textbf{Dependency on plug-in sensors}. The detection of a fine-grained face-screen distance value typically requires plug-in sensors, which may require an external battery to power the sensor.
  \item \textbf{Single-user mode}. Existing approaches only handle the single-user case, which may fail to take care of the user experience when multiple users are looking at the phone screen.
  \item \textbf{Detection cost}. The detection of face-screen distance may incur a high cost and impair the user experience, e.g., taking pictures using multiple cameras may shorten the battery life.
\end{enumerate}

To address these problems, we pose requirements for the design of a face-screen distance detection scheme on smartphone, accordingly. First, the detection scheme should tell the face-screen distance with a small latency and a high detection accuracy. Second, it should be independent of external sensors as well as external batteries. Third, it should take care of the multi-user mode. Finally, it is necessary to make the detection process transparent to the users, and minimize the use of phone camera or other built-in sensors to save power.

In this paper, we present a fine-grained detection method, called ``look into my eyes (LIME)'', that meets the above design requirements by minimizing the use of phone's front camera and inertial accelerometer to estimate the face-screen distance as well as to reduce the cost. Specifically, LIME captures the photo of the user's face only when the accelerometer detects that the specific motion of smartphone, and then estimates the face-screen distance by looking at the distance between the user's eyes. Besides, LIME is able to identify the face-screen distance when multiple users are facing the phone screen. The experimental results show that LIME can achieve a mean squared error smaller than 2.4 cm in all of our experimented scenarios. We implement an SMS application on Android phones with LIME integrated to enable the dynamic adjustment of font size by detecting the face-screen distance, e.g., the font size becomes larger when the detected face-screen distance increases.

The rest of this paper is organized as follows. We briefly review the related work in section \ref{sec:related_work}. We present the design of LIME in Section~\ref{sec:systemDesign}. We evaluate the performance of the LIME system with different metrics in Section~\ref{sec:evaluation}. Finally, we conclude the paper in section~\ref{sec:conclusion}.

\section{Related work}\label{sec:related_work}

\subsection{Fine-grained distance detection using smartphones}
Distance detection has been applied in many utility mobile applications, most of which utilize additional instruments to achieve fine-grained results. Distance detection systems based on the infrared~\cite{benet2002using,ponglangka2011eye} or ultrasonic~\cite{he2015optimizing,saad2011robust} leverage light-emitting devices, speakers and external sensors which are attached to the mobile phone. Other distance detection schemes depend on a camera together with auxiliary tools such as convex mirrors~\cite{murmu2014low}, or multiple cameras~\cite{mustafah2012stereo}.

LIME only makes use of the front camera, thereby minimizing the inconvenience caused by the external instruments. Meanwhile, LIME is different from those systems that are independent on external tools~\cite{konig2014new,shoani2015determining}: LIME employs an algorithm to determine when the photo-taking is needed to detect the face-screen distance. Camera takes images only when distance detection is necessary, while existing systems continuously measure the distance without ruling out those unnecessary situations that cause excessive energy cost.

\subsection{Camera based measurement}
Camera based method has been widely used in many mobile systems. Monocular camera has been used to measure the distance to the surface of objects \cite{yamaguti1997method}. In addition to measuring the distance, cameras are also applied to the speed measurement for vehicles in the traffic~\cite{schoepflin2003dynamic}. Besides the traditional measurement, the usage of camera is also extended to indoor localization. Powerful positioning system inside buildings can be implemented with the aid of camera~\cite{werner2011indoor}. The potential of camera is further exploited in indoor positioning system by reconstructing the indoor map with crowdsensed photos~\cite{gao2014jigsaw}.

However, most of the previous applications of camera in mobile systems fall in the scope of coarse granularity scenarios, which can even tolerant meter level error. LIME is different from these works in that we design a better solution to push usage of camera toward finer granularity applications, which is sensitive to decimetre and even centimeter level error. This can enable a richer measurement application for cameras.

\subsection{User experience optimization}
The most widely used distance detection application is the screen on/off legacy application that is installed on every smartphone operating system (the proximity sensor~\cite{proximitysensorAndroid} can determine the face-screen distance is small or not)~\cite{lane2010survey}. Similar applications can be found, such as the automatic screen-off upon the detection of the absence of the user~\cite{dalton2003sensing} for energy saving, dynamic resolution scaling~\cite{he2015optimizing} and frontal viewing angle~\cite{hu2013viri} for better visual experience. These applications depend on built-in sensors and are transparent to users, which implies the best user experience. LIME follows the same design requirement, and supports the optimization for the multi-users situation.

% into account, making the sharing experience also pleasant.

% More improvement in user experience optimization has been made by following researchers such as resolution scaling~\cite{he2015optimizing} and providing users frontal viewing experience from a slanted angle by photo processing~\cite{hu2013viri}.

\section{System Design}~\label{sec:systemDesign}
In this section, we present the design of LIME in details. First we describe the motivation of our design in Section~\ref{sec:motivation}. Then we present the architecture of LIME in Section~\ref{sec:overview}. We describe how LIME support single and multiple scenario in Section~\ref{sec:single} and Section~\ref{sec:multiple}. We also design an approach to reduce the power consumption of LIME in Section~\ref{sec:power}.

\subsection{Motivations}\label{sec:motivation}
% The detection of distance between human and mobile phones is important to improve the experience of interaction between user and smart phone. By measuring the distance, application can dynamically adjust screen on-off, screen resolution, screen luminance, font size and etc. But current approaches have two major problems.
In this part, we elaborate the main reasons that motivate the design of LIME: (1) most of existing face-screen detection methods involve auxiliary devices; (2) most of these approaches fail to consider the case when multiple users are facing the screen.

\textbf{Dependency on auxiliary devices}. The proximity sensors on today's smartphones can only determine whether an object is close to the phone screen or not. To detect the fine-grained distance between the object and phone, previous researchers refer to auxiliary devices (e.g. infrared sensors, ultrasonic sensors and multiple cameras). However, the auxiliary devices attached to phones will make the whole system inconvenient to carry and use, and they also introduce the overhead for communication with the phone as well as extra power consumption.

\textbf{Need for multi-user viewing}. People tend to share interesting things on mobile phone with their friends, which leads to many scenarios where multiple users view a single phone at the same time. Therefore, these application scenarios call for a scheme that can determine the appropriate distance to multiple users' faces and make dynamic adjustment on phone screen for balancing the viewing experience for them. However, most of the existing systems  can only determine the distance from screen to the closest user.

\subsection{System overview}\label{sec:overview}
To address the aforementioned problems, we propose a design of lightweight detection of face-screen distance, called ``look into my eyes (LIME)'', which has the following noteworthy features: (1) Independent of auxiliary devices, LIME computes the distance between users and phone screen by identifying the interpupillary distance (distance between users' eyes) in the photos taken by the front camera of phone; (2) it is designed to support good user experience for multiple users, since multiple users can be detected by photo; (3) we also exploit some optimization techniques to minimize the use of camera and motion sensors for reducing LIME's power consumption.
% detection between user and mobile phone. Designing such an approach face a couple of challenges. First, we can not use auxiliary devices, which means, all the hardware that we use should be built-in devices in mobile phone. Second, our approach should be able to detect the distance from the mobile phone to multiple users and take the user-experience of multiple users into consideration. Third, our approach should reduce the power consumption caused by distance detection as much as possible.
We implemented LIME into a legacy app on smartphones, i.e., SMS app, to dynamically change the font size based on the face-screen distance. Note that LIME can also easily be combined with other smartphone applications (e.g. browser, contacts).

\begin{figure}[htbp]
 \centering \includegraphics[width=3in]{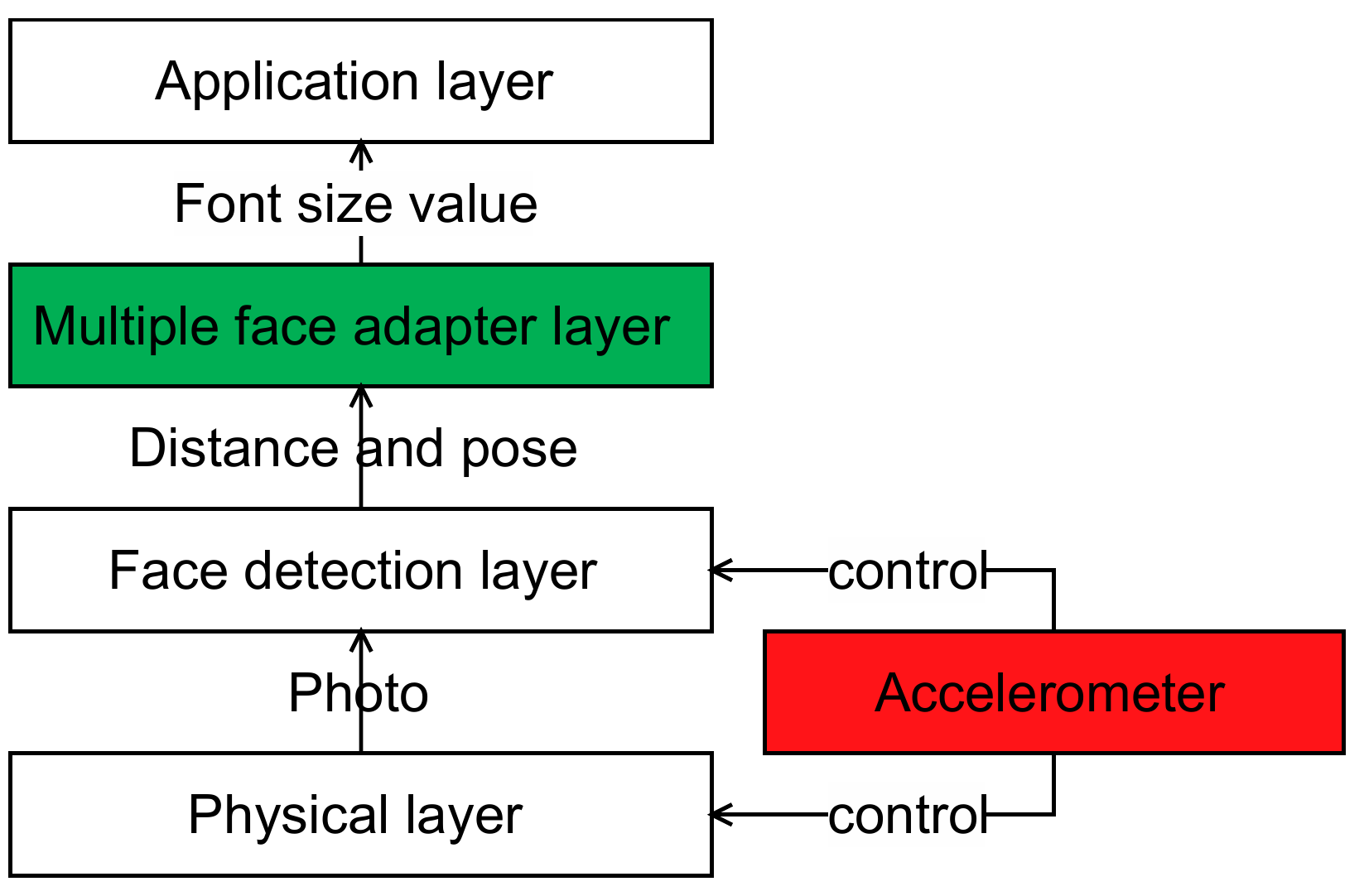} %\vspace{-0.1in}
\caption{System architecture of LIME.}
\label{fig:architecture}
\end{figure}

We show the four-layer architecture of LIME in Figure~\ref{fig:architecture}.
The first/bottom layer of LIME is the physical layer that calls the smartphone's front camera, which captures the photo of a user's or users faces and provides the photo to the upper layer.
The second layer of LIME is the face detection layer, which recognizes users' face from the photo and computes the position of the face. At this layer, we use Google face API~\cite{faceapi},
and the functions we have used in LIME are listed in Table~\ref{tab:table1}. The third layer of LIME contains the multi-face adapter and computes the appropriate
parameters based on the position of multiple faces in the photo. The top layer is the application layer which allows the controller of SMS application to
update the font size as soon as it receives the parameters from the third layer. In order to reduce the power consumption of LIME, we add a power controller
for the whole system, which can stop invoking the camera and face detection API when they are not in need.

\begin{table}[!b]
  \centering
  \begin{tabular}{|c|c|}
    \hline
    \multicolumn{2}{|c|}{\textbf{Google face API}} \\
    \hline
    \textbf{eyesDistance} & Returns the distance between the eyes. \\
    \hline
    \textbf{getMidPoint} & Sets the position of the mid-point between the eyes. \\
    \hline
    \textbf{pose} & Returns the face's pose. \\
    \hline
  \end{tabular}
  \caption{Face recognition functions.}
  \label{tab:table1}
\end{table}

\subsection{Single-user mode}\label{sec:single}
We first study the detection of face-screen distance in the single-user case.

\begin{figure}[htbp]
\centering
\includegraphics[width=3in]{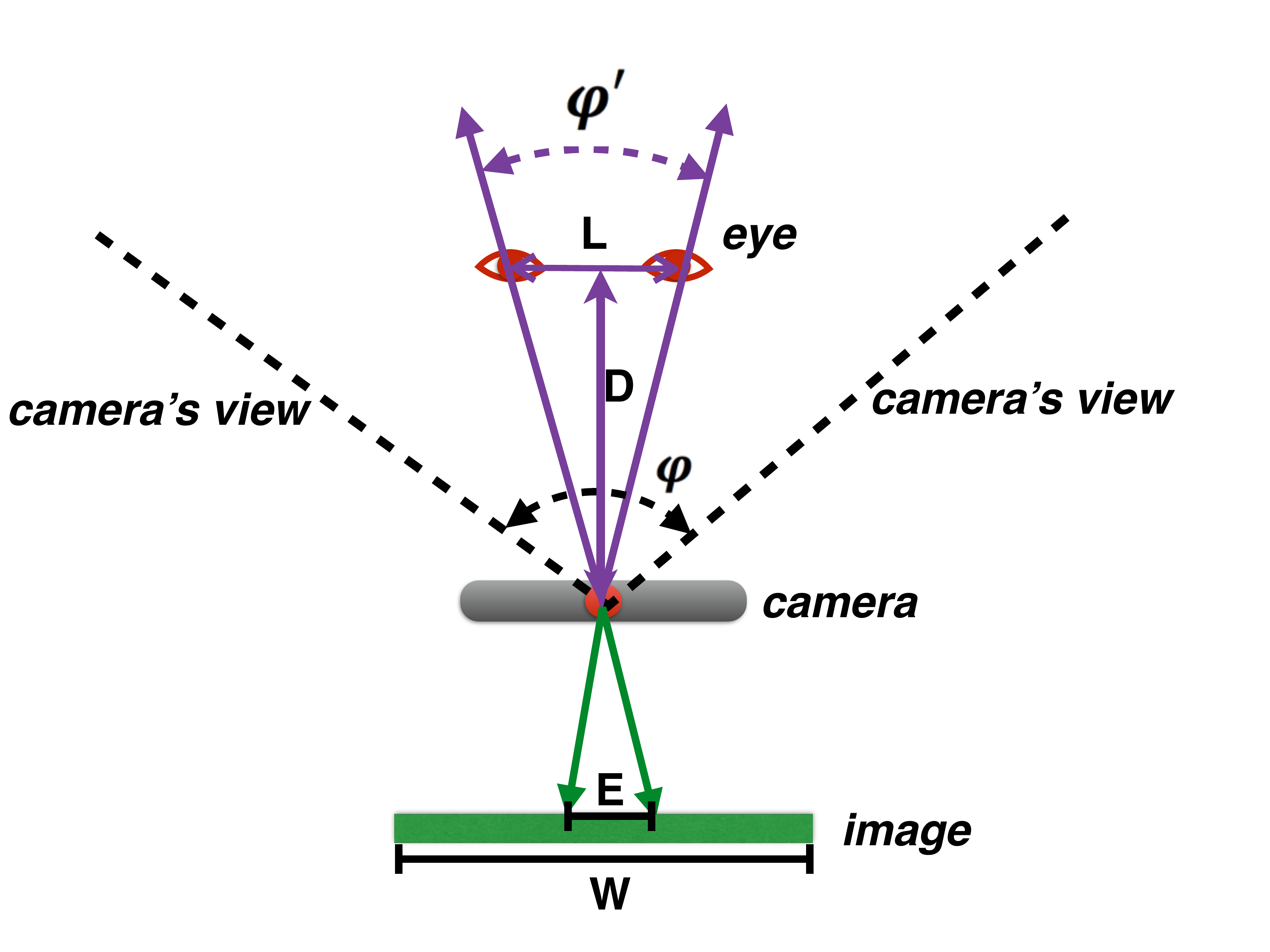}
\caption{Illustration of the distance estimation in the single-user case.} \label{fig:singleuser-2d}
\end{figure}

\textbf{Reference point and direction}. In this paper, we let the \emph{reference point} be the position of mobile phone camera and let the \emph{reference direction} be the direction which is perpendicular to the screen.
As Figure~\ref{fig:singleuser-2d} shows, we consider the user's position from a top-down 2D perspective, and the position of a user can be represented by a tuple $<D, \theta>$ in polar coordinates, where $D$ is the distance from the reference point and $\theta$ is the angle from a reference direction. Since we consider the scenario of single user, we only have to computing $D$ in this case.

\begin{figure}[htbp]
\centering
\includegraphics[width=3in]{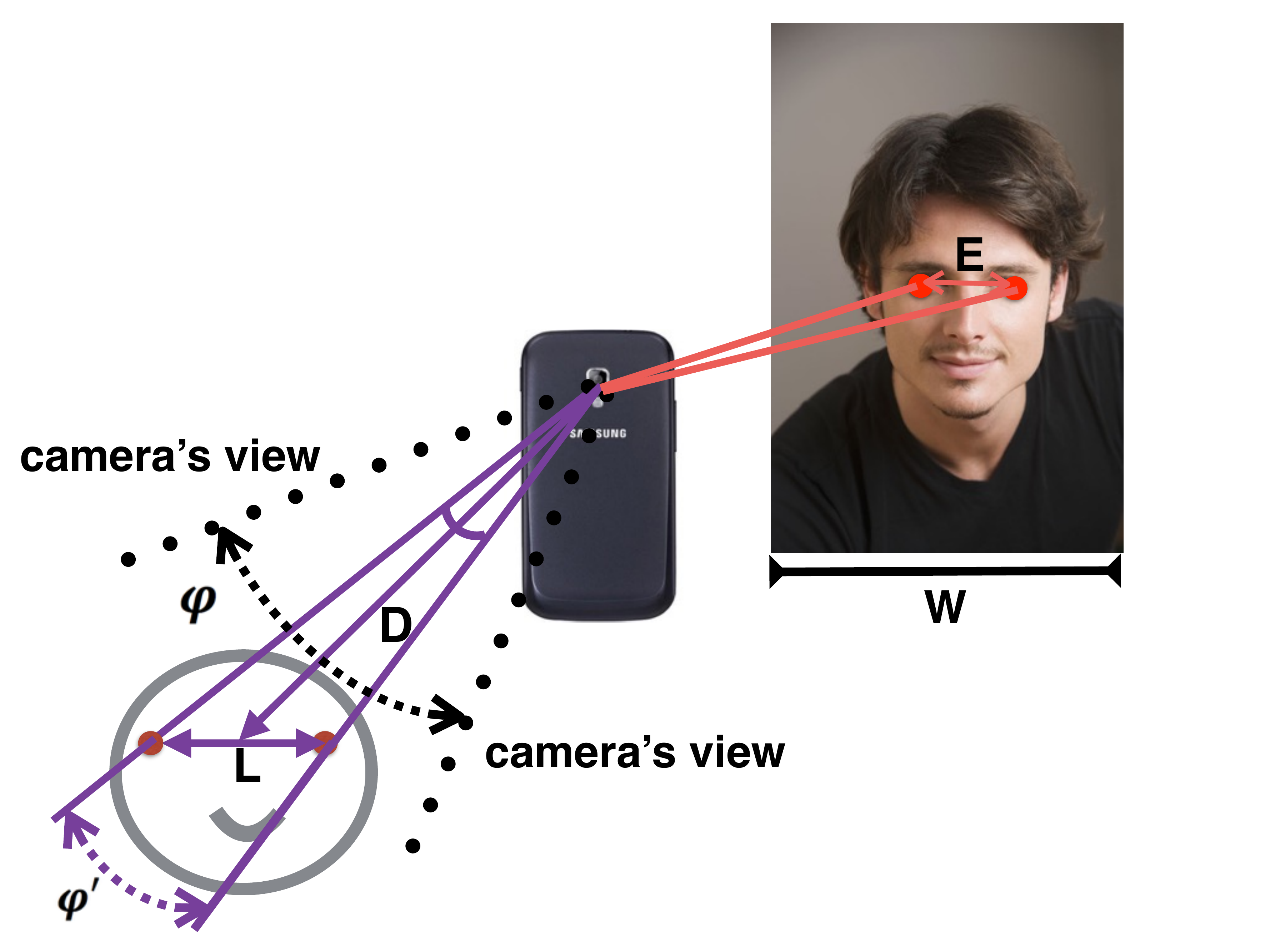}
\caption{Illustration of single-user case of LIME. } \label{fig:singleuser}
\end{figure}

\textbf{Exploiting the interpupillary distance}. As is known, the interpupillary distance (distance between users' eyes) of people varies little~\cite{Interpupillarydistance}, and it is feasible to leverage it to approximate the distance between user and phone. In Figure~\ref{fig:singleuser}, the phone can identify the interpupillary distance of the user in the photo, which helps compute the face-screen distance.

\textbf{Face-screen distance estimation}. Let $L$ denote interpupillary distance of people, $D$ denote the distance between user and mobile phone, $E$ denote interpupillary distance in the photo, and $W$ denote the width of photo. The width of an object in photo is determined by the width of this object in the camera view, and we have,
\begin{equation}
\label{eq:1}
\frac{\phi'}{\phi} = \frac{E}{W},
\end{equation}
where $\phi$ is the width of camera's view angle and $\phi'$ is the width of interpupillary distance in the camera's view angle.
From Figure~\ref{fig:singleuser-2d}, we can see that $\phi'$ can be determined by $D$ and $L$,
\begin{equation}
\label{eq:2}
\frac{\phi'}{2}=arctan\frac{L}{2D}.
\end{equation}
By combining (\ref{eq:1}) and (\ref{eq:2}), we have
\begin{equation}
\label{eq:3}
\frac{2arctan\frac{L}{2D}}{\phi}=\frac{E}{W}.
\end{equation}
By transforming (\ref{eq:3}), we can computes $D$ such as:
\begin{equation}
\label{eq:4}
D = \frac{L}{2tan\frac{\phi E}{2W}}.
\end{equation}

\subsection{Multi-user mode}\label{sec:multiple}
To extend LIME to support the multi-user case, we need to estimate the exact position of users, and figure out how to balance the viewing experience for multiple users. Since the position of a single user is described by a tuple $<D, \theta>$, for multiple users, we should not only know the distance $D$, but estimate the angle $\theta$. We have described how to determine $D$, and now we show how to compute $\theta$ in this part.

\begin{figure}[htbp]
\centering
\includegraphics[width=3in]{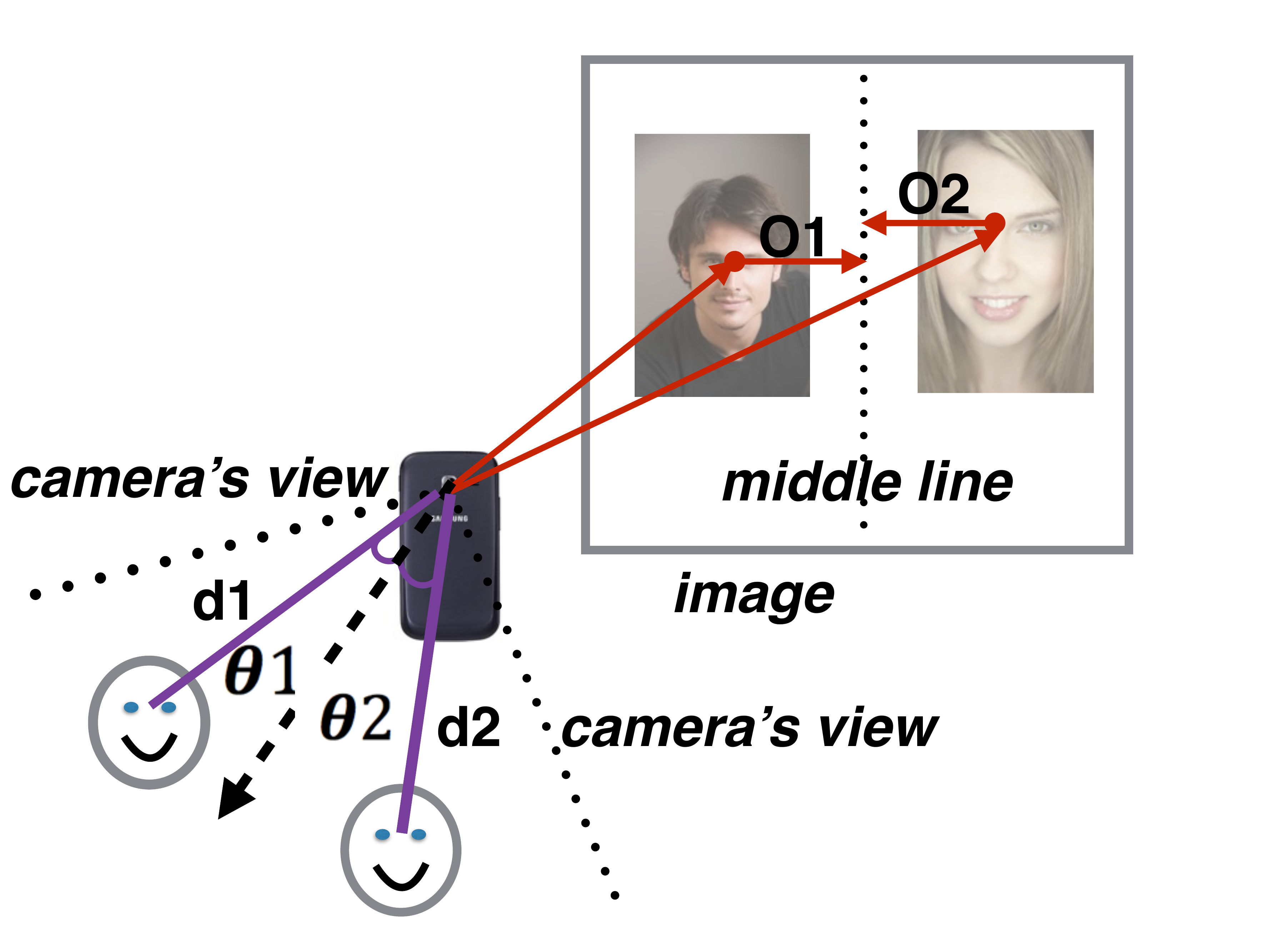}
\caption{Illustration of multi-user case of LIME.} \label{fig:multipleuser}
\end{figure}

\begin{figure}[htbp]
\centering
\includegraphics[width=3in]{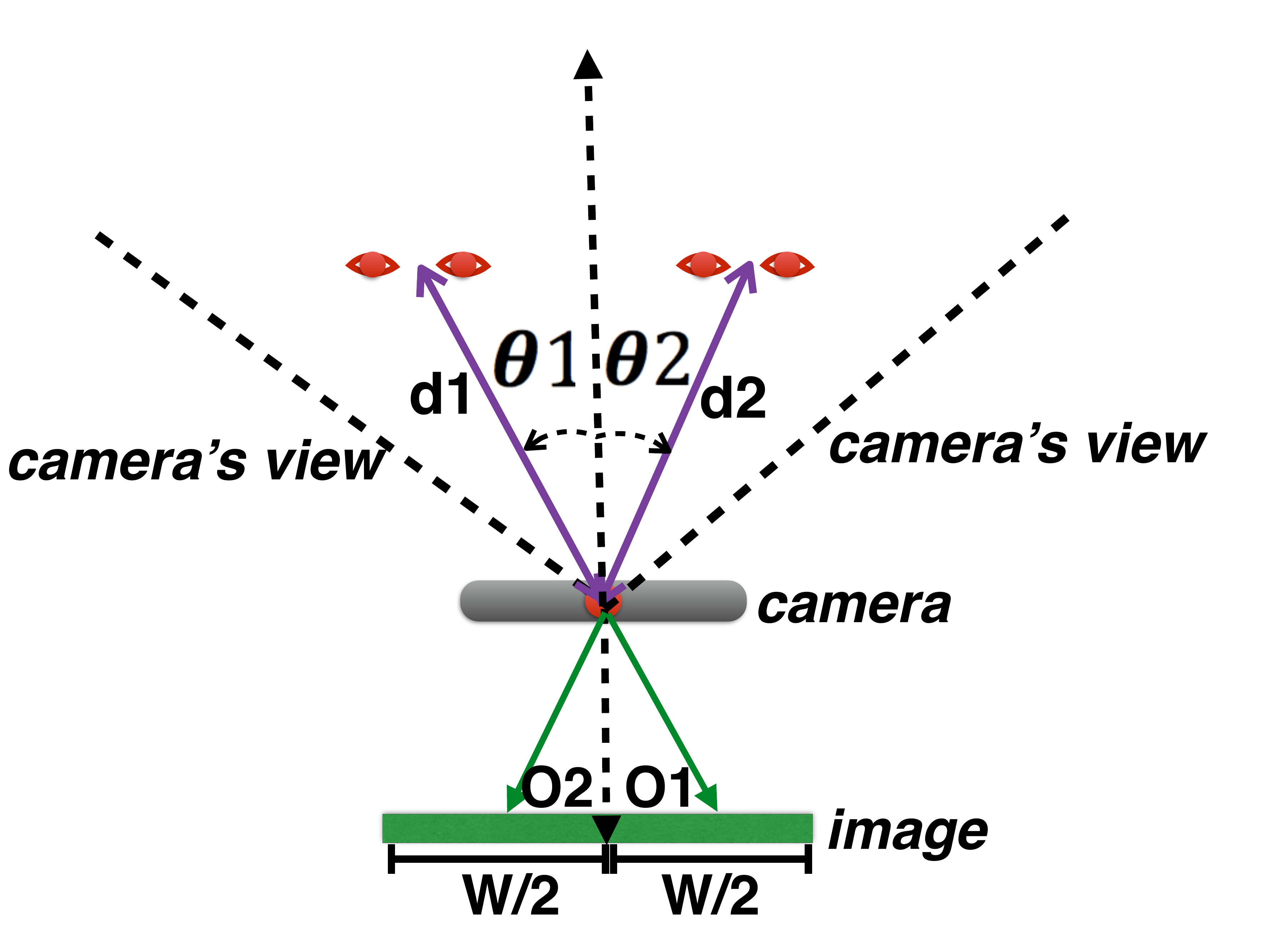}
\caption{Illustration of the distance estimation in the multi-user case.} \label{fig:multipleuser-2d}
\end{figure}

\textbf{Estimation of the angle from the reference direction}. Figure~\ref{fig:multipleuser} shows an example scenario of how LIME computes $\theta$ for two users, and the method can be easily extended to the case of more than two users. Figure~\ref{fig:multipleuser-2d} is a 2D schematic diagram from top-down perspective, from which we have
\begin{equation}
\label{eq:5}
\frac{\theta_i}{\phi} = \frac{O_i}{W},
\end{equation}
where $O_i$ ($i=1,2$ in this example to represent two users' IDs) is the distance from the center of user~$i$' two eyes in the photo to the middle line of the photo and $\theta_i$ is the angle from the reference direction for user~$i$. Then we can get the formula which computes $\theta$ by transforming (\ref{eq:5}).
\begin{equation}
\label{eq:6}
\theta_i = \frac{\phi O_i}{W}.
\end{equation}

%\begin{figure}[t]
%\begin{minipage}[t]{0.45\linewidth}%
%
%
%\centering
%\includegraphics[width=1.5in]{pic2.pdf}
%\caption{Illustration of multi-user case of LIME.} \label{fig:multipleuser}
%
%
%\end{minipage}
%\hfill%
%\begin{minipage}[t]{0.45\linewidth}%
%
%\centering
%\includegraphics[width=1.5in]{pic4.pdf}
%\caption{Illustration of the distance estimation in the multi-user case.} \label{fig:multipleuser-2d}
%
%\end{minipage}
%\end{figure}

%\begin{figure}[ht]
% %
%\begin{minipage}[t]{0.45\linewidth}%
%\centering
%\includegraphics[width=1.6in]{pic2.pdf}
%\caption{Illustration of multi-user case of LIME. }\label{fig:multipleuser}
%\end{minipage}
%\hfill%
%\begin{minipage}[t]{0.45\linewidth}%
%\centering
%\includegraphics[width=1.6in]{pic4.pdf}
%\caption{Illustration of the distance estimation in the multi-user case. } \label{fig:multipleuser-2d}
%\end{minipage}
%\end{figure}

\textbf{Dynamic font size in the demo app given multiple users}. Next, we show how to combine LIME with the demo SMS app by dynamically adjusting the font size, according to the location of multiple users, where LIME follows the rule of balancing the user-experience of all the users. Specifically, LIME tries to find a ``center'' of all the users and adjust the font size according to the position of ``center''. The position of ``center'' should minimize the quadratic sum of distance between the ``center'' and the position of all the users. We can get the position of the ``center'' by solving the optimization problem below:
\begin{align*}
\text{Minimize}
~~ \sum_{i=1}^nDist(i,c)^2,
\end{align*}
where $n$ is total number of people and $Dist(i,c)$ is the distance between the position of user~$i$ and position of ``center''. We use the tuple $<D_i,\theta_i>$ to denote the position of user~$i$ and the tuple $<D_c,\theta_c>$ to denote the position of ``center'' in polar coordinates. So the distance can be computed by
\begin{equation}
\label{eq:7}
Dist(i,c) = \sqrt{D_i^2+D_c^2-2D_1D_ccos(\theta_i-\theta_c)}.
\end{equation}
Put (\ref{eq:7}) into our optimization goal and we have
\begin{equation}
\label{eq:8}
\sum_{i=1}^nD_i^2+D_c^2-2D_1D_ccos(\theta_i-\theta_c).
\end{equation}
The optimization goal can be rewritten as
\begin{equation}
\label{eq:9}
\sum_{i=1}^n(D_icos\theta_i-D_ccos\theta_c)^2+(D_isin\theta_i-D_csin\theta_c)^2.
\end{equation}

Let $p_i=D_icos\theta_i$, $p_c=D_icos\theta_c$, $q_i=D_isin\theta_i$ and $q_c=D_isin\theta_c$. We can get
\begin{equation}
\label{eq:10}
\sum_{i=1}^n((p_i-p_c)^2+(q_i-q_c)^2).
\end{equation}
% Note that $p_c$ and $q_c$ are in different part of the formula, so we can compute the minimal value of different part of this equation separately.
By computing the derivative of different parts of the formula above, we derive the minimal condition: $p_c = \sum_{i=1}^n\frac{p_i}{n}$ and $q_c = \sum_{i=1}^n\frac{q_i}{n}$. Then we can compute $D_c$ and $\theta_c$ from $p_c$ and $q_c$.
\begin{equation}
\label{eq:11}
D_c = \sqrt{(\sum_{i=1}^n\frac{D_icos\theta_i}{n})^2+(\sum_{i=1}^n\frac{D_isin\theta_i}{n})^2}.
\end{equation}
\begin{equation}
\label{eq:12}
\theta_c = arctan\frac{\sum_{i=1}^n\frac{D_isin\theta_i}{n}}{\sum_{i=1}^n\frac{D_icos\theta_i}{n}}.
\end{equation}

\subsection{Optimizing the power consumption}\label{sec:power}
In LIME, the use of front camera of smartphone will raise a concern regarding the power consumption. High power consumption leads to a short battery life, which impairs the user experience.
% In order to design a system with good user-experience, we still need to optimize the power of LIME.
Note that most of the power consumption of LIME is caused by the front camera. So we can save the battery life by minimizing the use of the camera. Instead of the continuous detection the face-screen distance, we only need to detect the distance when the user changes the position of his phone---we use the built-in accelerometer to detect the user's motion, and LIME only opens the camera to take a photo upon the detection of the phone movement; it closes the camera when the phone remains relatively stationary.

\section{Evaluation}~\label{sec:evaluation}
In this section, we evaluate the performance of LIME under four criteria via experiments: detection accuracy, response latency, power consumption, and user experience.
\subsection{Implementation}
We have implemented our LIME prototype system on Google Nexus 6 and XiaoMi note running Android 5.0 OS and a xiaomi-customized OS based Android 4.4 respectively. The LIME is
incorporated into an SMS application on smartphones, which enables dynamic adjustment of font size based on the detected screen-face distance.

\subsection{Detection accuracy}
\begin{figure}[htbp]
 \centering \includegraphics[width=2.2in]{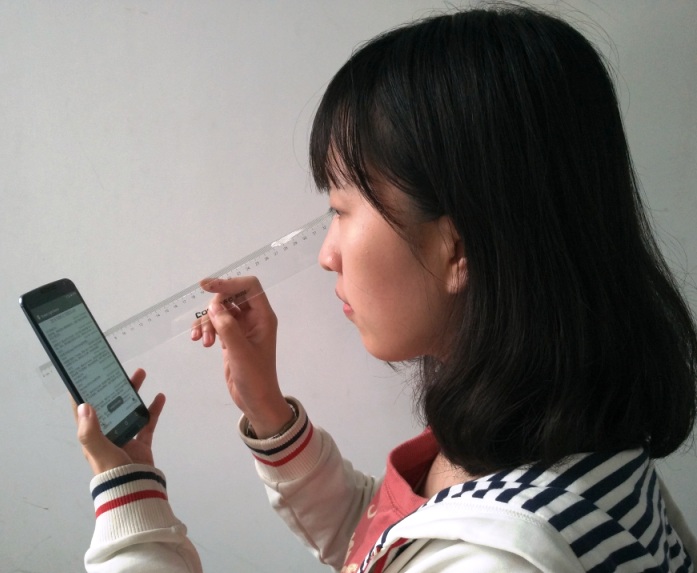} %\vspace{-0.1in}
\caption{Participant holds the phone and the ruler at the same time, and keeps the phone and the ruler placed orthogonally. We record the LIME-detected distance on phone and the actual distance read from the ruler.}
\label{fig:demo_measure}
\end{figure}

\begin{figure}[htbp]
\centering
\includegraphics[width=3in]{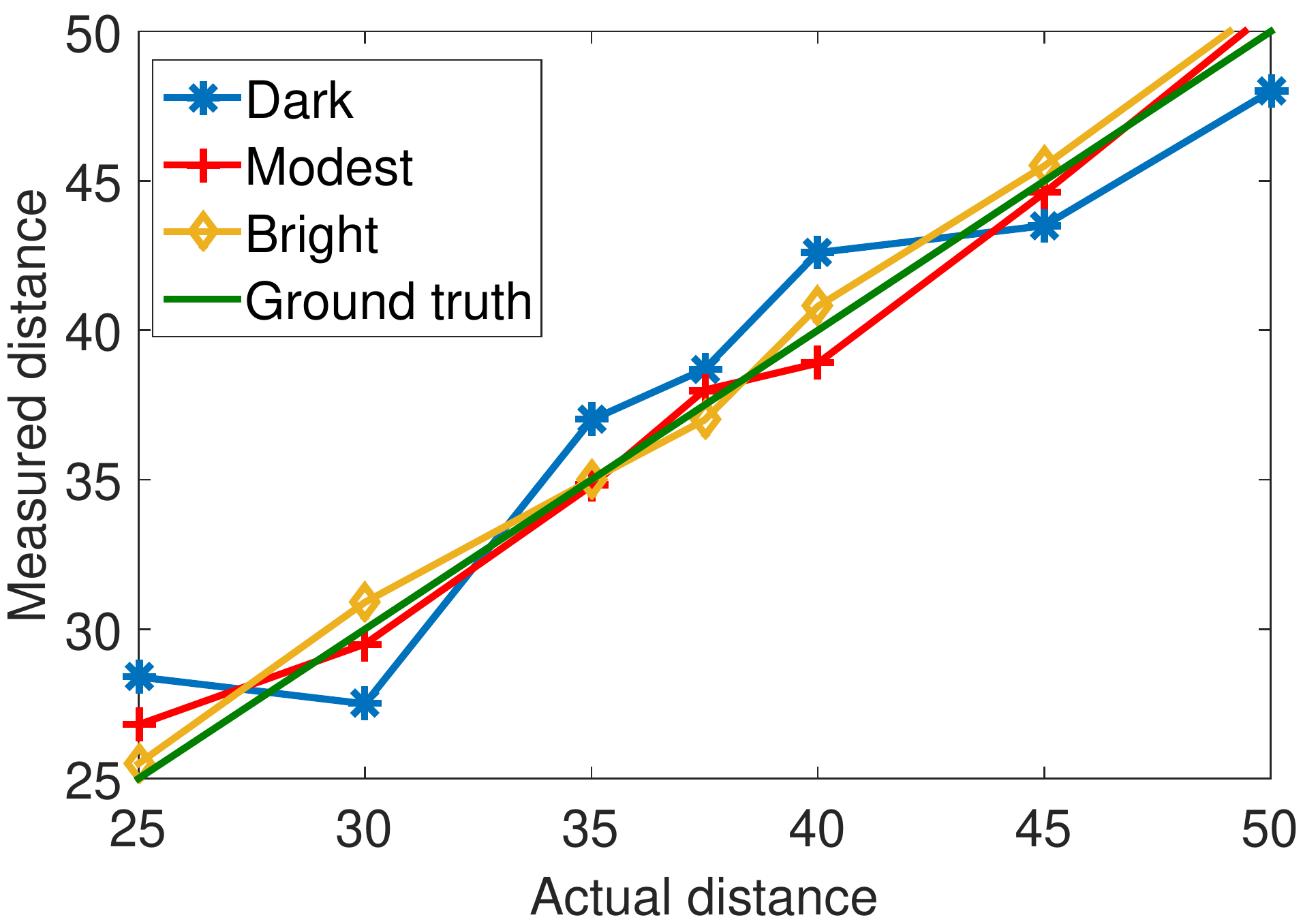}
\caption{The detection accuracy in different light conditions.} \label{fig:accuracy}
\end{figure}

% To assure the optimal font size adjustment, it first requires LIME to acquire accurate distance measurement results.
In the detection accuracy test, we consider different lighting conditions, invite 10 participants to hold the phone,
and measure the actual distance using a ruler (shown in Figure~\ref{fig:demo_measure}) to compare with the detection results.
We measure each distance by five times and show the average value of them. The results are shown in Figure~\ref{fig:accuracy},
the ``dark'', ``modest'' and ``bright''curves show the results of indoor environment with dime light, indoor environment with modest light,
and outdoor environment with bright light at noon.

The ``ground truth'' curve shows the distance measured by the ruler, which is supposed to be a linear line to indicate how far other curves deviates from the ground truth. As we can observe from the results, in all of the three lighting conditions, the distances detected by LIME are linearly correlated with the ground truth curve, and the mean squared errors under these circumstances are 0.6761, 0.8976, and 2.3878 respectively, which leads to the conclusion that LIME has an excellent detection performance both indoor and outdoor.

\begin{figure}[htbp]
\centering
\includegraphics[width=3in]{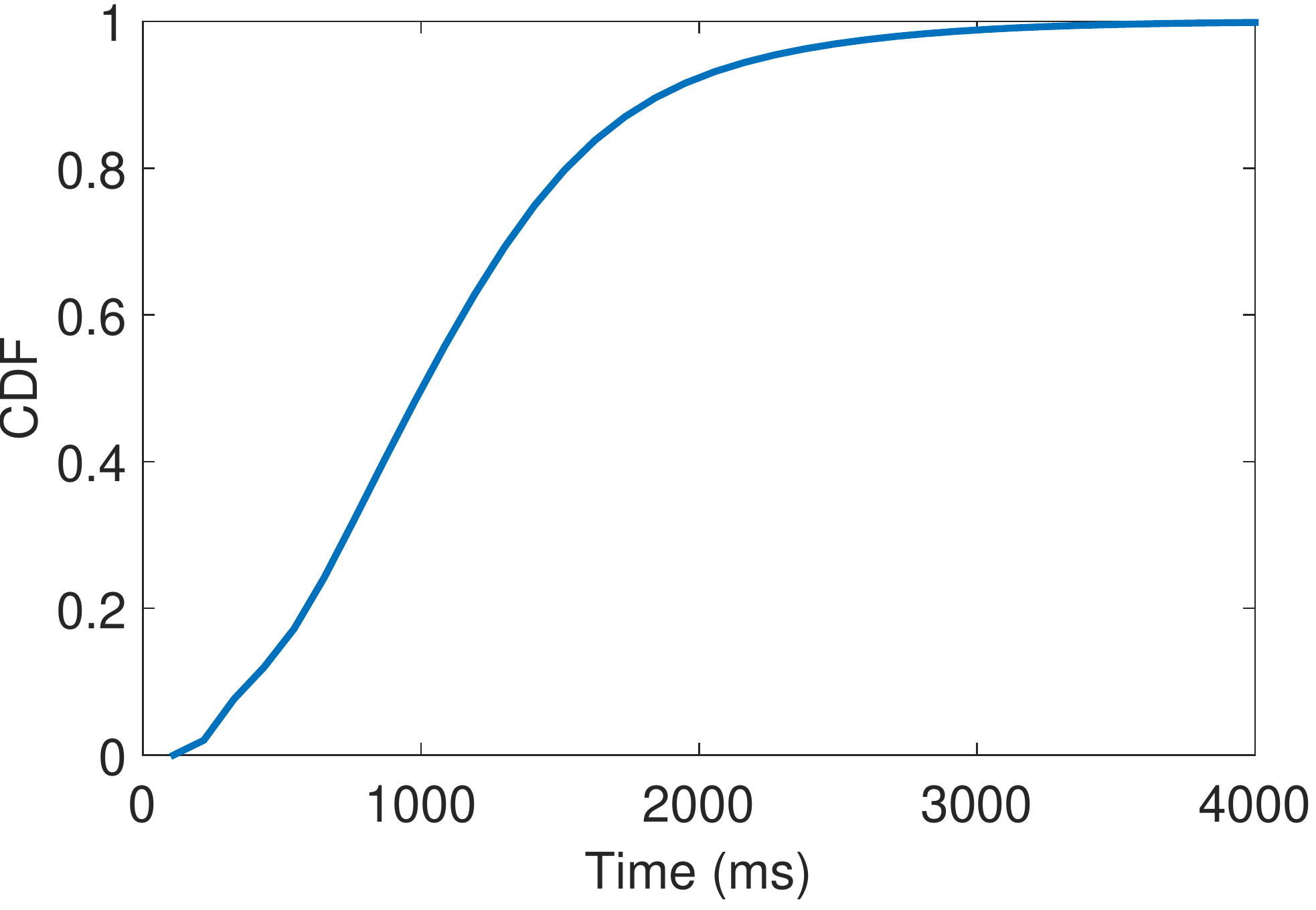}
\caption{CDF of the response latency.} \label{fig:response-latency}
\end{figure}

%\begin{figure}[ht]
%\begin{minipage}[t]{0.45\linewidth}%
%
%\centering
%\includegraphics[width=1.5in]{accuracy.pdf}
%\caption{The detection accuracy in different light conditions.} \label{fig:accuracy}
%\end{minipage}
%\hfill%
%\begin{minipage}[t]{0.45\linewidth}%
%\centering
%\includegraphics[width=1.5in]{latency.pdf}
%\caption{CDF of the response latency.} \label{fig:response-latency}
%\end{minipage}
%\end{figure}

\subsection{Response latency}
A small response latency makes it possible for LIME to be applied to a wide range of mobile applications. We require 10 participants to change the
face-screen distance as fast as possible so that we could omit the time of moving the phone. Then we calculate the response latency using the timer
that also shows the latency in the screen. To fully evaluate the latency of LIME, we conducted the experiments for 30 times so that we could collect the
extreme latency data.

The cumulative distribution function (CDF) is shown in Figure~\ref{fig:response-latency}. As we can see, the probability that the response latency smaller
than 2~sec is about 0.9, which means that in most cases LIME has a small latency.

% with a seldom happens. One millisecond response latency is acceptable by consumers range according to our common sense.

\subsection{Power consumption}
As pointed out previously, LIME is energy-efficient as it only measures the distances by using the front camera to capture the photo when the phone is moved. In this section, we evaluate the power consumption of LIME.

\begin{figure}[htbp]
\centering
\includegraphics[width=3in]{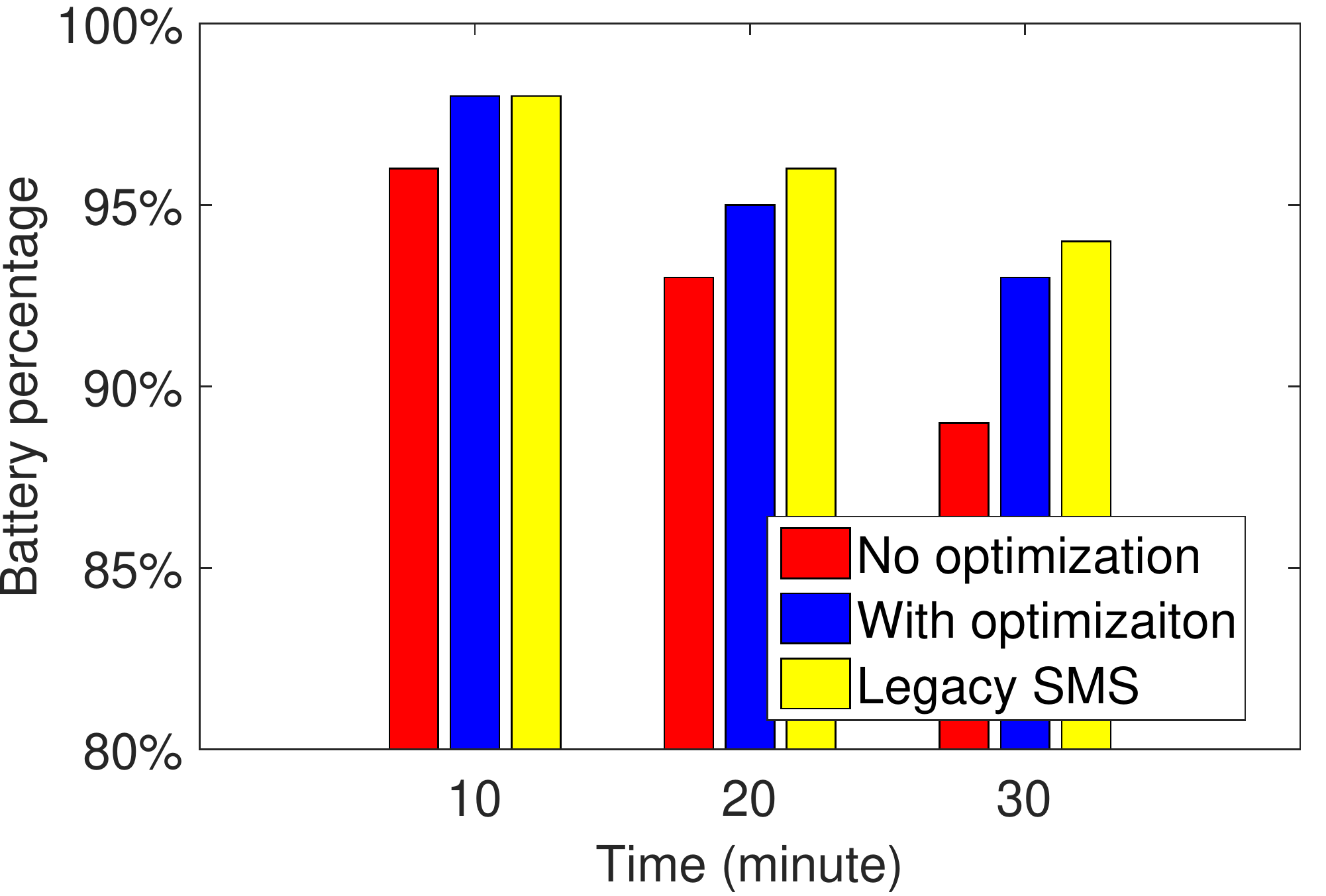}
\caption{The energy consumption result.} \label{fig:power-consumption}
\end{figure}

We turn the phone into the flight mode so that we could rule out other unnecessary power consumption. We compare the average power consumption for different time durations, namely 10, 20, 30 minutes, under three scenarios: (1) legacy SMS app without font size adjustment; (2) SMS app with continuous use of the front camera for font size adjustment; and (3) SMS app with LIME (on-demand use of the front camera for font size adjustment). The results of the three cases are labeled as ``Legacy SMS'', ``No optimization'' and ``With optimization'' in Figure~\ref{fig:power-consumption}.

The results clearly show that the optimization could save the energy consumption by 40 percent compared with the design without optimization algorithm. What is more interesting is that the energy consumption of LIME is comparable to that of the legacy SMS app, which means that most of energy consumption is caused by the phone screen instead of LIME. Thanks to the optimization that rules out unnecessary photo-taking operations conditions, LIME achieves an excellent performance in maintaining battery life.

\begin{figure}[htbp]
\centering
\includegraphics[width=3in]{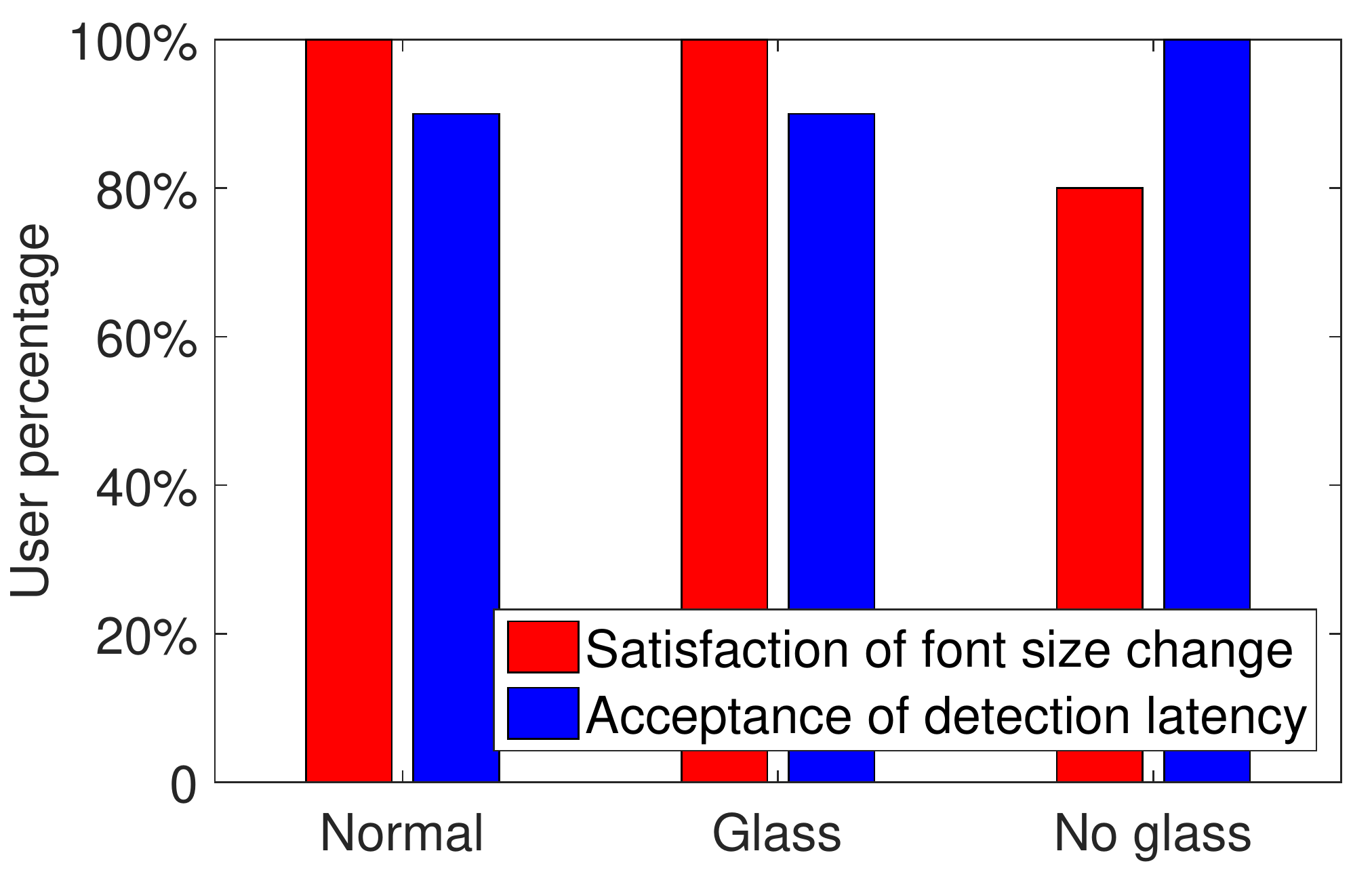}
\caption{The percentage of people who are satisfied with font size change and response latency.} \label{fig:user-study}
\end{figure}

%\begin{figure}[ht]
%\begin{minipage}[t]{0.45\linewidth}%
%\centering
%\includegraphics[width=1.5in]{power.pdf}
%\caption{The energy consumption result.} \label{fig:power-consumption}
%\end{minipage}
%\hfill%
%\begin{minipage}[t]{0.45\linewidth}%
%\centering
%\includegraphics[width=1.5in]{user_study1.pdf}
%\caption{The percentage of people who are satisfied with font size change and response latency.} \label{fig:user-study}
%\end{minipage}
%\end{figure}

\subsection{User Experience Study}

\subsubsection{Single user case}
In our first study of user experience, we invite 30 participants (13 males and 17 females, ranging from 18 to 51 years old) from a university in China to use our system for one week. After one week experiment, we collect the feedback from these participants. These participants on average spent 3.2 hours on their smartphones per day. In order to differentiate the effect of participants' eyesight, we intentionally allocate them into three different groups based on their eye sight capabilities, as well as whether they are wearing glasses.
There are altogether three groups: people whose eye sights are normal; people who have myopia without glasses; people who have myopia but wear glasses. Each group contains 10 participants.

We collect experiment data in two aspects: (1) the satisfaction of the font size adjustment and (2) the tolerance of response latency. The number of people who are satisfied with the font adjustment and the detection latency of each group is shown in Figure~\ref{fig:user-study}. The results show that participators whose eye sights are normal or who wear glasses are all comfortable with the font size change, at the same time, most of them claim that LIME makes reading experience of short messages more pleasant and relaxing. However, some of participants who are not wearing glasses and suffering from myopia may claim that although LIME greatly improves the reading experience of short messages, it would be better that the font size could be bigger. As for the tolerance of response latency, most of people in all these three groups think that the latency does not impair the usability.

%\begin{table}[!b]
%  \centering
%  \begin{tabular}{|c|c|c|c|}
%    \hline
%    \multicolumn{4}{|c|}{\textbf{Satisfaction rate}} \\
%    \hline
%     &healthy &myopia w/ glasses &myopia w/o glasses \\
%    \hline
%    healthy                 &10  &8.5 &7.5 \\
%    \hline
%    myopia w/     &8.5  &9.5 &8 \\
%    \hline
%    myopia w/o  &7.5  &8   &7.5 \\
%    \hline
%  \end{tabular}
%  \caption{Satisfaction rate for two users using LIME at the same time.}
%  \label{tab:table2}
%\end{table}

\begin{table}[!b]
  \centering
  \begin{tabular}{|@{}c|c|c|c|}
    \hline
    \multicolumn{4}{|c|}{\textbf{Satisfaction rate for font size.}} \\
    \hline
     \diagbox[trim=l]{User1}{User2}&Norm &Myopia w/ glasses &Myopia w/o glasses \\
    \hline
    Norm                 &100\%  &85\% &75\% \\
    \hline
    Myopia w/ glasses    &85\%  &95\% &80\% \\
    \hline
    Myopia w/o glasses &75\%  &80\%   &75\% \\
    \hline
  \end{tabular}
  \caption{Satisfaction rate for the two-user case.}
  \label{tab:table2}
\end{table}

\subsubsection{Multi-user case}
In the second study, we invite pairs of participants to use the SMS app with LIME, taking all different combinations of eye sights into consideration. Since the multi-user study mainly focuses on whether LIME provides consumers with the comfortable font size, we ask them to give their opinions on our system whether they are satisfied with the font size change, and the results are shown in Table~\ref{tab:table2}.

% after using LIME with a full mark of 10, and the average scores is shown in Table~\ref{tab:table2}.

As we could see from the above results, in general, the participators speak highly of our system and most of them admit that LIME makes reading of short messages easier. At the same time, we can notice that the people who have similar eye sights always have better user experience than those who are in poor eye sight conditions. However, we cannot deny the fact that the poor-eye-sight people obviously have better reading experience than before.
%since they also give us a relatively high remarks.
We believe that taking multi-user case into consideration improves the usability of LIME and expands the applicability of the face-screen detection technique.

\section{Conclusions}~\label{sec:conclusion}
In this work, we present a fine-grained detection method, called ``look into my eyes (LIME)'', which merely utilizes the front camera and inertial accelerometer of the smartphone to estimate the face-screen distance. To save battery life, LIME captures the photo of the user's face only when the accelerometer detects that the motion of mobile phones. LIME estimates the face-screen distance by looking at the distance between the user's eyes, which is a viable approach for either single-user or multi-user case. The experiments show that LIME can achieve a mean squared error smaller than 2.3878 cm in all of our experiment scenarios, and it incurs a small overhead on battery life when combined with an SMS application that enables dynamic adjustment of font size according to the detected face-screen distance.

\section{Acknowledgements}~\label{sec:acknowledgements}
We thank the anonymous reviewers and shepherd for their comments and feedback to help us improve our work. We are grateful to
the members of the Mobile group in Peking university for their insightful comments and suggestions. We also appreciate the all the volunteers for their participation in our experiments and offering valuable feedback. This research is funded by National Natural Science Foundation of China (NSFC) under grant number 61572051 and 61401169.

% Or, just remove \balance and give up on balancing the last page.
%
\balance

% If you want to use smaller typesetting for the reference list,
% uncomment the following line:
% \small
\bibliographystyle{abbrv}
\bibliography{reference-list}

\end{document}